\newcommand{\beq}{\begin{equation}}
\newcommand{\enq}{\end{equation}}
\newcommand{\beqarr}{\begin{eqnarray}}
\newcommand{\enqarr}{\end{eqnarray}}
\newcommand{\secref}[1]{Sec.\ \ref{#1}}
\newcommand{\figref}[1]{Fig.\ \ref{#1}}

\newcommand{\bk}{{\bf k}}

\newcommand{\bq}{{\bf q}}
\newcommand{\bp}{{\bf p}}

\documentclass{iopart}
\usepackage{iopams}
\usepackage{graphicx,subfigure}
\begin{document}

\title{Models of coherent exciton condensation}

\author{P B Littlewood$^\dagger$, P R Eastham, J M J Keeling,
F M Marchetti, B D Simons and M H Szymanska} 

\address{Theory of Condensed Matter, Cavendish Laboratory, Cambridge
CB3 0HE, United Kingdom.}
\address{and $^\dagger$ also at National High
Magnetic Field Laboratory, Pulsed Field Facility, LANL, Los Alamos NM
87545}

\date{\today}  
\begin{abstract}
That excitons in solids might condense into a phase-coherent ground
state was proposed about 40 years ago, and has been attracting
experimental and theoretical attention ever since. Although
experimental confirmation has been hard to come by, the concepts
released by this phenomenon have been widely influential. This
tutorial review discusses general aspects of the theory of exciton and
polariton condensates, focussing on the reasons for coherence in the
ground state wavefunction, the BCS to Bose crossover(s) for excitons
and for polaritons, and the relationship of the coherent condensates
to standard lasers.
\end{abstract}

\pacs{71. 35. Lk, 71. 36. +c}


\section{Introduction}
An electron and a hole optically excited within a solid are oppositely
charged, and bind together to form a bosonic exciton. Since the mass
of this particle is typically small, there has long been interest in
the possibility of obtaining a Bose-Einstein condensate (BEC) at
cryogenic
temperatures\cite{Moskalenko,Blatt,Keldysh-Kopaev}. Experimentally
this has proved challenging, because excitons are not the ground state
of the system, and a cold equilibrium gas needs to be prepared on a
shorter time scale than the excitons can decay. There have been many
approaches to this problem (for some reviews see
\cite{BEC_review1,BEC_review2,Hanamura-Haug}) with particularly
important systems being $\mathrm{Cu_2O}$ \cite{CuOexpt} (where dipole- and
spin-forbidden transitions are harnessed to produce excitons with long
lifetimes) $\mathrm{CuCl}$ \cite{CuClexpt} (which has very stable biexcitons)
and two-dimensional coupled quantum wells (where electrons and holes
are physically separated by a tunnel
barrier)\cite{bilayer,Kash91,Butov02a,Butov02b,Snoke02}.

In a situation where the dominant decay mechanism is by dipole
radiation, the opportunity to confine the light inside an optical
microcavity\cite{Microcavity} allows one instead to work with the
coupled eigenstates of the electron-photon problem, namely
polaritons\cite{Hopfield}.  The polariton effective mass can be made
much lighter -- as small as $10^{-5}$ of the electron mass -- and a
naive estimate of the critical temperature thus even higher. One now
has an extra handle on the experimental system because coupling of the
cavity to an external electromagnetic field allows both coherent and
incoherent pumping of the system. Some recent work has demonstrated
the onset of stimulated emission\cite{LeSiDang,Savvidis00}, parametric
oscillation in a driven cavity\cite{Baumberg00} as well as the
development of spontaneous\cite{Deng02,Deng03,Weihs03} optical
coherence in semiconductor microcavities.

This article will review some of the theoretical aspects of the
exciton problem, particularly associated with the construction of an
appropriate wavefunction for a condensate that is based on bound pairs
of fermions. The basic insight on this problem was provided by the
work of Keldysh and collaborators\cite{Keldysh-Kopaev,Keldysh-Koslov}
using a variational wavefunction in close analogy to the BCS
wavefunction for superconductivity. This approach was extended by
Nozi\`eres and Comte\cite{Nozieres-Comte} who showed how such a
wavefunction smoothly interpolates between the regime of a dilute Bose
gas and a dense two-component plasma, and then reworked for
superconductors to provide a theory of the BCS-BEC
crossover\cite{Nozieres-Schmitt-Rink,RanderiaReview}. We will explain
here how the collective mode spectrum changes qualitatively between
the two limits, and connect this spectrum to the familiar picture of a
dilute Bose gas.

In the polariton condensate, the pairs of fermions can resonantly
decay into photons, so the order parameter is shared between the two
coherent degrees of freedom -- the photon electric field and the
excitonic polarisation. Inspection of this physical system also
reminds us that BEC in an interacting system leads to a broken
symmetry corresponding to phase coherence of the dipole oscillators --
so that in a broad sense BEC of polaritons makes a kind of laser. To
make that relationship explicit, we shall discuss the Dicke model of
localised dipole-active transitions coupled to a cavity field. It
turns out that a straightforward generalisation of the BCS-like
wavefunction for the coupled system provides a good description of the
problem\cite{Eastham00,Eastham01}. This mean field theory can be
extended\cite{Keeling04} to discuss the analog of the BCS-BEC
crossover -- which in this case connects the limits of BEC of a dilute
gas of polaritons with a higher density system where the coherence is
produced through the self-consistent optical field (as in a laser). It
turns out that the density scale for this crossover corresponds to a
separation between excitons(i.e. excitations of the localised
transitions) which is the geometric mean of two parameters: the
wavelength of photons at energies of the order of the polariton
splitting, and the spatial separation of the localised
transitions. The former length scale is generally a few tenths of a
micron, while the latter is greater than the exciton Bohr radius. Thus
the density range where the correct description of the problem is
polaritonic BEC is probably quite limited.

The model we use for polaritonic condensation is similar to that
describing other systems based on (quantum) oscillators coupled by
resonance with a bosonic field\cite{Eastham03}: two prominent examples
are arrays of small Josephson junctions coupled in a microwave
cavity\cite{Barbara99,Harbaugh00}, and cold fermionic atoms coupled to
a molecular Feshbach resonance\cite{Timmermans01,Holland01}. The
phase-coherent ground state describing the excitonic insulator can be
mapped to the coupled bilayer quantum Hall state near $\nu = 1/2$
\cite{EisensteinBilayer,MacDonaldBilayer}.

We shall stress that the important issue associated with excitonic or
polaritonic condensation is coherence, rather than momentum condensation as in
the weakly interacting Bose gas. Because we are dealing with physical systems
that are open, and can exchange excitation with the environment, the coherence
in the system may be destroyed not only thermally by internal excitations
(i.e. particle-hole pairs or phase fluctuations) but also by coupling to
external baths (which may be non-thermal). These decoherence processes can
produce crossovers to other, more familiar, coherent phenomena such as lasing,
before driving the system into complete
incoherence\cite{Szymanska02,Szymanska03,Marchetti04}.

\section{Possible phases of the electron-hole system}
The Hamiltonian of the electron-hole system consists of the kinetic
energy of the separate components, and the Coulomb interaction between
them. Written in second-quantised notation, 
\begin{equation} \label{hbasic}
H = H_o + H_{coul}, \end{equation} where
          \begin{equation}
          \label{H_o}
          H_o = \sum_{k} \left [ \epsilon_{ck}a_{c, k}^{\dagger}a_{c, k}
               +  \epsilon_{vk}a_{v, k}^{\dagger}a_{v, k} \right ] ,
          \end{equation}
     and
          \begin{equation}
          \label{H_coul}
               H_{coul} = \frac{1}{2} \sum_{q} \left [ V^{ee}_q \rho^e_q \rho^e_{-q} +
               V^{hh}_q \rho^h_q \rho^h_{-q}
               - 2 V^{eh}_q \rho^e_q \rho^h_{-q} \right ].
          \end{equation}
     $a_{c, k}^{\dagger}$ and $a_{v, k}^{\dagger}$ are creation operators
     for electrons in the conduction and valence bands. The density operators
          are $\rho^e_q = \sum_k a_{c, k+q }^{\dagger}a_{c, k}$,
          $\rho^h_q = \sum_k a_{v, k}a_{v, k+q }^{\dagger}$.
          $V(q)$ is the Coulomb interaction, and for a homogeneous three-dimensional system
          $V^{ee}=V^{hh}=V^{eh} = 4 \pi /\epsilon q^2$.
          It is interesting also to consider the two dimensional situation of separate layers of
          electrons and holes,
          where $ V^{ee}_q = V^{hh}_q = 2 \pi / \epsilon q$,
          $V^{eh}_q = 2 \pi e^{-qd} /\epsilon q$, and $d$ is the interlayer separation.
          For parabolic bands, then
          $\epsilon_c(k) = \hbar^2 k^2/2m_e$;
          $\epsilon_v(k) = -E_g - \hbar^2 k^2 / 2 m_h$.

     The natural units are the exciton Rydberg,
     $Ry^*=\frac{\mu e^4}{2 \epsilon^2 \hbar^2} = \frac{\mu}{m}\frac{1}{\epsilon^2} Ry$, and the
     exciton Bohr radius,
     $a^* = \frac{\epsilon \hbar^2}{\mu e^2} = \epsilon \frac{m}{\mu} a_o$.
     Here $Ry = 13.6 \; eV$ is the
     Hydrogenic Rydberg, $\mu$ the reduced mass, and $a_o = 5 \times 10^{-10} \mathrm{m}$ the hydrogenic Bohr radius. One of the
     principal reasons that semiconductor systems are so interesting is that a combination of
     small band mass $\mu$ and large dielectric constant $\epsilon$ means that $a^*$ can often be very large -- so even at moderate
     excitation levels, the characteristic separation between excitons can be made comparable to their Bohr radius.
     It is convenient to measure the density $n$ (of electron-hole pairs) of the system in units of the Bohr radius
     by defining the dimensionless
     parameter $r_s$:  in three dimensions $\frac{1}{n}= \frac{4\pi}{3}(r_sa^*)^3$, and in two dimensions
     $\frac{1}{n} =  \pi (r_s a^*)^2$.

This is not the complete Hamiltonian for electrons and holes in a real
solid with a real bandstructure that includes all the effects of Bloch
electrons. The model is a good approximation for semiconductors with a
light mass and a large dielectric constant because the effective Bohr
radius is much longer than the physical lattice constant. Most
importantly for our purposes, this Hamiltonian separately conserves
the number of electrons and the number of holes. Interband tunnelling
and interband exchange is neglected here. This neglect is not
quantitatively important for determining the ground state, but if
present will break the conservation of electrons and holes and
formally prohibit a superfluid ground state\cite{Guseinov,PBL96}.

The electron-hole system is surely one of the simplest model systems
in condensed matter physics. The ground state(s) of this model are
likely to include various kinds of quantum solids and
liquids\cite{PBL03}. The relevant parameters are the density (measured
by $r_s$), the mass ratio of electron to hole $\Gamma = m_e/m_h$, and,
for 2D bilayers, the separation $d$. If $\Gamma \gg 1$, then we are
discussing hydrogen, where we expect that the two basic phases are
either a molecular solid of $H_2$, or at very high densities a
metallic crystal --- where the electrons delocalise. (There may of
course be solid phases with different crystal structures within each
of these basic types). The pressures required to obtain this are
immense. There is no regime where a gas of individual excitons is
expected.

The molecular stability of $H_2$ is large --- the heat of formation of
a molecule from two atoms of hydrogen is roughly 1/3 $Ry$ --- which is
why the phase diagram at moderate to low densities $r_s > 1$ should be
dominated by solid phases in a system with high hole mass.  In
contrast, with electron and hole of roughly equal mass the binding
energy of the biexciton $X_2$ --- the analogue of $H_2$ --- is about
one order of magnitude smaller, and the biexcitonic molecule is
corresponding large. In consequence, the biexcitonic solid (nearly
equal masses) is expected to form (if at all) only at low densities
($r_s \approx 5-10$). At higher densities, it is plausible to believe
that this solid will melt to form a fluid phase. The form of this
quantum fluid is easy to imagine at very high densities $r_s \ll 1$,
because here the kinetic energy of both species (scaling like
$r_s^{-2}$) will overcome the Coulomb binding (scaling as $r_s^{-1}$)
and a good description would be of two interpenetrating Fermi
liquids. At lower density, there will be fluctuations leading to the
transient appearance of excitonic atoms and molecules in the solid,
and these configurations will preponderate at larger $r_s$.

There is also the possibility of an exciton crystal, which would be an
atomic (Mott) insulator in contrast to the biexcitonic band
insulator. Such a phase should be readily stabilised in 2D bilayer
systems at large $d$ and small $r_s$, where it is more easily
recognised as two coupled Wigner crystals. The 2D bilayers should also
have reduced stability of the biexciton (because of dipole repulsion
between two excitons) and so are more likely to support quantum fluid
phases over a wider range of density that 3D systems.

This model is amenable to study by advanced numerical techniques,
including variational\cite{Zhu96}, quantum\cite{Senatore} and
path-integral\cite{Ceperley} Monte Carlo methods. However, the full
phase diagram has not yet been established theoretically.

\section{Theory of the excitonic insulator}
From now on we shall be concerned entirely with the fluid phase, and
immediately the question arises about whether it is condensed. There
are three major aspects to the character of a condensate: the
statistical physics of bosons (the conventional texbook view of BEC);
phase coherence of the order parameter; and superfluidity.

Since at low density, $r_s \gg 1$, we have a fluid that can be sensibly
thought of as atomic, one expects Bose-Einstein condensation
(BEC). Conventionally, one regards BEC as a phenomenon associated with
the statistical physics of weakly interacting bosons. While this may
be appropriate for a very dilute gas of strongly bound atoms, it is
less clear whether this is the appropriate physical description of a
dense two-component plasma. So the first issue is how to write down a
wavefunction in terms of the fermionic components, that nevertheless
recognisably describes bosons in the dilute limit.

Such a wavefunction must contain within it the important physical
characteristic of macroscopic phase-coherence. Phase coherence is a
consequence of interactions, but even infinitesimally small
interactions in boson systems convert the (highly degenerate) ground
state obtained by considering the statistical physics of BEC to a
robust phase-locked condensate. It turns out that in exciton and
polariton systems the phase coherence has physical consequences for
the interaction with electromagnetic radiation that are different from
in a superconductor and quite characteristic of the condensed state.

The third, and most subtle issue, is that of superfluidity. In an
extended fluid with Galilean invariance, continuous changes in the
superfluid phase generate supercurrents that can flow without
dissipation. Whether or not the exciton condensate is a true
superfluid (or instead a density wave) and what in fact would be the
correct superfluid response is a subtle topic that is not yet
completely resolved.

Before addressing excitonic systems, it is useful to start with a
brief review of BEC in the dilute Bose gas (for a general and complete
exposition, see e.g. the book by Pethick and
Smith\cite{PethickBook}). Since the first observation of BEC in cold
atomic gases in 1995, there has of course been tremendous activity in
this field that we will not attempt to review. Our discussion will be
focussed on the effect of interactions and coherence.

\subsection{Coherence and interactions in atomic BEC}

BEC as a phenomenon in statistical physics is usually presented in
terms of the occupancy of single particle states $n_q$, indexed by
momentum $q$. For a free particle of mass $M$, the states are occupied
according to the Bose factor
\begin{equation}
n_q = \frac{1}{e^{\beta (E_q - \mu )} -1}
\end{equation}
where $E_q = \hbar^2 q^2/2M$ is the kinetic energy of the boson, and
$\beta = 1/k_BT$.  The total number of particles in the system is then
fixed by
\begin{equation}
\label{eq:NBEC}
N = \sum_q n_q = \int dE  \frac{D(E)}{e^{\beta (E-\mu)} -1} \;\;\;,
\end{equation}
which is actually an equation determining the chemical potential as a
function of temperature. Here $D(E) \propto E^{d/2-1}$ is the density
of states in dimension $d$.

As temperature is lowered the Bose factor in Eq. (\ref{eq:NBEC})
becomes sharply peaked in the vicinity of the chemical potential ---
and in consequence $\mu$ must increase so as to allow the integral to
conserve $N$. Remarkably, in dimensions $d>2$ it turns out that the
integral remains finite even as $\mu \to 0$, and therefore the
chemical potential reaches the bottom of the band at a non-zero
temperature $T_{BEC}$. By dimensional arguments it is clear that this
temperature is close to the degeneracy temperature $k_BT_o =
\hbar^2n^{2/d}/m $, where the thermal de Broglie wavelength reaches
the interparticle separation. Below this temperature $\mu$ remains
clamped to the bottom of the band and the state with zero momentum has
an occupation proportional to the total number of particles $N$ in the
system.

\subsection{Interactions, broken symmetry and collective modes in the dilute atomic condensate}
\label{sec:collective}

This picture is not an inaccurate way to describe a dilute gas of
weakly interacting bosons, but it misses a crucial feature of BEC ---
macroscopic phase coherence, and the {\em rigidity} of the
condensate\cite{AndersonRMP,NozieresBEC}. If we have a system of
macroscopic size $\Omega = L^d$ then, as L is very large, there is
only a small separation in energy, $\propto L^{-2}$, between the
ground state $q_0=0$ and the low-lying excited states with momenta $q$
of order $1/L$. So while the number state $|N, q_0 \rangle = N^{-1/2}
(c_0^\dagger)^N \left | 0 \right >$ is indeed lowest in energy, states
of the form $(c_0^\dagger)^{N-m_1-m_2 ...}  (c_1^\dagger)^{m_1}
(c_2^\dagger)^{m_2} ... \left | 0 \right > $ have an energy that is
greater only by an amount of order $(m_1 + m_2 + ...)/L^2$, provided
we restrict ourselves to momenta of order $1/L$. (We use $c_k$ as the
annihilation operator for a boson in momentum state $q_k$ and
$\hat\phi({\bf r})= \sum_k <{\bf r}|k>c_k = \Omega^{-1/2}\sum_k
\exp(i\bk \cdot {\bf r} ) c_k $ for the field operator.)

What breaks this near degeneracy are interactions between particles.  Consider
the (bosonic) Hamiltonian $H_0+H_{int}$, for particles of mass $m$ in an
external potential $V_{ext}$. We have \beq
\label{eq:Hoboson}
H_0 = \int d{\bf r}  \hat\phi^\dagger({\bf r} ) \left [ -\frac{\hbar^2 \nabla^2}{2m} + V_{ext} ({\bf r} ) \right] \hat\phi ({\bf r} ),
\enq
together with the interaction term
\begin{equation}
\label{eq:Hintboson}
H_{int} = 
\frac{1}{2}\int d{\bf r} d{\bf r^\prime} V({\bf r}-{\bf r^\prime}) \hat\phi^\dagger ({\bf r}) \hat\phi^\dagger ({\bf r^\prime}) \hat\phi ({\bf r^\prime}) \hat\phi({\bf r}).
\end{equation}
(Often this interaction is modelled by a short range term $V({\bf r} -
{\bf r^\prime}) = V_0 \Omega \delta ({\bf r} -{\bf r^\prime})$, an approximation which is
sensible once the thermal de Broglie wavelength $\lambda_T = (2 \pi
\hbar^2 \beta / m)^{1/2}$ is much larger than the interparticle
spacing --- or equivalently that $T \ll T_o$.)

We can discuss the effect of the interaction energy using an
appropriate trial wavefunction.  Rather than the number states $|N,
q_0\rangle $ we instead consider coherent states \beq
\label{eq:cohstate}
|\Lambda, q_0 \rangle = e^{\lambda c_0^\dag} e^{-|\lambda|^2/2}
|0\rangle.
\enq
This wavefunction is a state of well-defined phase, with an
expectation value of the number of particles of $ \langle N \rangle = |\lambda|^2 $: 
\beq
\langle \Lambda | \hat\phi ({\bf r} ) | \Lambda \rangle = \lambda= 
|\lambda|e^{i\theta}.
\enq
 The phase is conjugate to the number of particles
     since we can generate a number state $\left |N\right>$ as follows:
     \begin{equation}
     \label{numberstate}
     \int d\theta \; e^{-iN\theta} \left|\Lambda , q_0 \right> \propto
     \left| N , q_0 \right > .
     \end{equation}

To show how the interactions make the system resistant to fragmentation,
consider a mixed state
\beq \label{eq:cohmixed}
\left | \Psi \right> = e^{-\lambda^2/2} e ^{\lambda (\cos(\alpha)c_0^\dagger + \sin(\alpha)c_1^\dagger )} \left | 0 \right >.
\enq This state has a population fragmented between two different momenta: $N_0 = \lambda^2 \cos^2 (\alpha )$, 
$N_1 = \lambda^2 \sin^2 (\alpha )$, $N_0 +N_1 = N$
(here we restrict $\alpha, \lambda$ to be real without loss of generality).
The interaction energy can be straightforwardly evaluated
\beq
\label{eq:mixing}
\left < \Psi \right| H_{int} \left| \Psi \right >
= \frac{1}{2}\lambda^4  V_o [1 + \frac{1}{2}\sin^2(2\alpha)] = \frac{1}{2} V_o N^2 + V_o N_0 N_1.
\enq
Since $V_o$ is positive (repulsive interactions) the energy is clearly
minimised by the pure state with $\alpha=0$ or $\alpha = \pi/2$; which
of these two is lowest is determined by the kinetic energy. Notice
that this answer does not depend on the momenta of the two states (as
long as they are both small).  The interaction energy provides an
extensive energy penalty for any mixture, as long as the interactions
are repulsive.

The coherent states (\ref{eq:cohstate}) and (\ref{eq:cohmixed}) are
often described as ``breaking global gauge symmetry'', in that they
have a well-defined overall phase. This feature, however, is not
essential for the arguments above. We could have reached an identical
conclusion using number states, which do not have an overall phase,
because to leading order in $N$ the energy of (\ref{eq:cohstate}) is
identical to that of $|N,q_0\rangle$, and the energy of
(\ref{eq:cohmixed}) is identical to that of $|N_0,q_0; N_1,q_1
\rangle$. The point is that a single state in an interacting system
has a particular phase relationship between different components of
the wavefunction. In a condensate, the energy differences between
states with different phase relationships can be large, even when the
matrix elements are small, because statistics ensures that some modes
become macroscopically occupied. Thus phase \emph{relationships} which
in the normal state are washed out by thermal fluctuations, or by
applied fields, become robust in the condensed state.

The generally accepted definition of a Bose condensate is as a system
with off-diagonal long-range order\cite{HuangBEC}. This means that the
one-body density matrix, $\langle
\hat\phi^\dagger({\bf r})\hat\phi({\bf r}^\prime)\rangle$, approaches a non-zero
constant for large separations $|{\bf r}-{\bf r}^\prime|$. The practical
upshot of this is that one can see interference effects between
particles removed from widely separated regions of the condensate, so
that off-diagonal long-range order is indeed connected to the presence
of unusual phase relationships in the wavefunction. Interestingly,
interactions in condensates should enforce phase relationships
involving more than two removed particles\cite{NozieresBEC}, although
the presence of such higher-order coherence is not required by the
definition of off-diagonal long-range order. Note also that standard
wavefunctions, such as (\ref{eq:cohstate}), often contain higher
orders of coherence than required for the presence of off-diagonal
long-range order.

To understand the collective behaviour of a condensate we need to
introduce an order parameter for condensation. One way to do this is
to define the order parameter from the one-body density matrix
according to $\langle \hat\phi^\dagger({\bf r})\hat\phi({\bf r}^\prime)\rangle
=\phi^\ast({\bf r})\phi({\bf r}^\prime)$. This defines the order parameter
$\phi({\bf r})$, which is a complex classical field called the condensate
wavefunction.  The Ginzburg-Landau free energy for the condensate
wavefunction $\phi({\bf r})$ is
\beq
\label{eq:GL}
F[\phi] = \int d{\bf r} \left [ \frac{\hbar^2}{2m} \left| \nabla \phi ({\bf r}
) \right | ^2 + (V_{ext} ({\bf r} ) - \mu ) |\phi ({\bf r} ) | ^2 +
\frac{V_o}{2} \left | \phi ({\bf r} ) \right | ^4 \right ] \enq The formal
route to this functional constructs an action based on the model of
interacting bosons above, from which the G-L theory emerges as a
classical saddle point (see, e.g.\cite{Stoof99}). The path from here
on is discussed in many textbooks\cite{PethickBook}, and we will just
quote results.

If we minimise the free energy of Eq. (\ref{eq:GL}) we obtain an equation
for the ground state wavefunction $\phi_o$ which is the
Gross-Pitaevski equation \beq
\label{eq:GP}
\left [ - \frac{\hbar^2}{2m} \nabla^2 + V_{ext} ({\bf r} ) - \mu +
V_o\left | \phi_o ({\bf r} ) \right | ^2 \right ] \phi_o({\bf r} ) =
0.  \enq If we now consider small deviations $\phi = \phi_0 + \eta$,
then we can determine the energy of quadratic fluctuations: \beqarr
\fl \int d{\bf r} \left ( \begin{array}{c c} \eta^* & \eta \end{array}
\right ) \left ( \begin{array}{c c} -\frac{\hbar^2}{2m} \nabla^2 +
V_{ext}-\mu + 2 V_o |\phi_o|^2 & + V_o \phi_o^2 \\ +V_o\phi_o^{*2} &
-\frac{\hbar^2}{2m} \nabla^2 + V_{ext}-\mu + 2 V_o |\phi_o|^2
\end{array} \right ) \nonumber \\ \times \left ( \begin{array}{c} \eta
\\ \eta^* \end{array} \right ). \label{eq:GL2} \enqarr The
fluctuations mix the real and imaginary components of the fields: what
is happening is simplest to envisage for a uniform condensate
($V_{ext} = 0$); then the solution of Eq. (\ref{eq:GP}) determines the
chemical potential $\mu=V_0|\phi_o|^2 = V_o n_o$, and after taking a
Fourier transformation the matrix at the core of Eq. (\ref{eq:GL2})
becomes \beq \left ( \begin{array}{c c} \epsilon_k + V_o |\phi_o|^2 &
V_o \phi_o^2 \\ V_o\phi_o^{*2} & \epsilon_k + V_o |\phi_o|^2
\end{array} \right ), \enq where $\epsilon_k = \hbar^2 k^2 / 2
m$. Since we have a coupling between $\eta$ and $\eta^*$, not only is
the normal average $<\eta \eta^*>$ non-zero, but also the anomalous
average $<\eta \eta>$. Note that when we determine the dynamics of the
new wavefunctions, i.e. turning Eq. (\ref{eq:GL2}) into a
Schr\"odinger equation, we need to get the time dependence straight by
looking for solutions of the form $(\eta^\ast \; \; \eta)=(\eta_0^\ast
e^{i\omega_k t} \; \; \eta_0 e^{-i\omega_k t})$. This leads to an
eigenvalue spectrum determined by \beq \left (
\begin{array}{c c} -\omega_k + \epsilon_k + V_o |\phi_o|^2 & V_o
\phi_o^2 \\ V_o\phi_o^{*2} & +\omega_k + \epsilon_k + V_o |\phi_o|^2
\end{array} \right ). \enq The new excitation modes of the condensate
thus have the dispersion first derived by Bogoliubov \beq
\label{eq:bogoliubov}
\omega_k = \sqrt{ \epsilon_k^2 + 2 V_o n_o \epsilon_k}.  \enq This
spectrum is acoustic in the long-wavelength limit $k \xi \ll 1$, where
$\xi = \hbar/(2mn_oV_o)^{1/2}$ is the healing length.  One may also
easily check that in the long wavelength limit this mode descibes
fluctuations of the phase of the order parameter, as we expected.

This approach connects the microscopic theory to the insight of Landau that a fluid with only phonons as the low energy excitation spectrum cannot absorb arbitrarily small amounts of energy whilst also conserving momentum. The coherence in the underlying wavefunction generated an acoustic spectrum, and that produces superfluidity.

\subsection{Mean field theory for excitons}
Now we return to the consideration of exciton systems, and our first concern is to write down an analogous wavefunction for BEC, when our bosons consist of bound pairs of fermions.

The wavefunction for a single exciton is just a wavepacket of electron-hole pairs, viz.
\begin{equation}
     \label{excitonwfn}
     \left | \Phi_q \right > = \sum_k \phi (k,q) a_{c,k+q}^\dagger a_{v,k} \left| 0 \right >,
     \end{equation}
Here our vacuum state $|0\rangle$ is a filled valence band and empty conduction band; consequently
$a_{v,k}$ creates a valence band hole. Eq. (\ref{excitonwfn}) describes an exciton with centre of mass momentum $q$, and $\phi(k,0)$ is thus just the Fourier transform of the real space exciton wavefunction in relative coordinates.
This is manifestly not a boson, but let us write a coherent state in analogy to
Eq. (\ref{eq:cohstate}) as follows:
\begin{equation}
     \label{eq:coherentexciton}
 |\Psi_{MF}> =    e^{\lambda \sum_k \phi (k,0) a_{c,k}^\dagger a_{v,k}}\left| 0 \right>.
     \end{equation}
Writing a wavefunction with fermion operators in the exponential is not necessary, because unlike bosons, we cannot have two fermions in the same state. So we can manipulate this wavefunction into something more familiar. We generalise the hydrogenic state to a variational function $g(k)$ and then expand the exponential, noting that the series terminates after the second term:
\begin{eqnarray}
|\Psi_{MF}> & = & \prod_{\vec k} e^{g(k) a_{c,k}^\dagger a_{v,k}}\left| 0 \right> \nonumber \\
&=&\prod_{\vec k} [ u_{\vec k} + v_{\vec k} a_{c, k}^{\dagger}
                                                   a_{v, k} ]|0\rangle.
\label{eq:keldyshwfn}
\end{eqnarray}
In the last line we have written $g(k)=v(k)/u(k)$ and have normalised the wavefunction so that $|u_{ k}|^2 + |v_{ k}|^2 = 1$.
$v(k)$ may now be taken as a variational function, and this wavefunction was written down by Keldysh and Kopaev\cite{Keldysh-Kopaev} in complete analogy to the BCS theory of superconductivity.

Provided $v_k$ (in general complex) has the same phase for all
momenta this is a coherent state in the same sense as the bosonic
state. But this wavefunction is in general richer than for bosons, as
it has an explicitly fermionic description and a variational function
$v_k$.

\subsection{BCS to BEC crossover for excitons}
\label{sec:BCS-BEC}
The variational functions $u(k)$ and $v(k)$ should be evaluated by minimising the expectation value of the Coulomb Hamiltonian, Eq. (\ref{hbasic}). The details have been discussed in many places and for many different geometries, for example by
\cite{Nozieres-Comte,PBL96}, and we will just review the main
results. Just as in a BCS model of superconductivity, we have an order
parameter corresponding to the broken gauge symmetry (phase coherence),
and a gap in the excitation spectrum.

In order to control the density, we introduce the chemical potential
$\mu$ for the introduction of electron-hole pairs with density $n$. We
then minimize the free-energy 
\begin{equation}
\label{eq:chemical}
F  = < H_o + H_{Coul}> - \mu <n>,
\end{equation}
with respect to the variational parameters $v_{k}$.
Setting $\partial F / \partial v_{\vec k} = 0$ and considering only
$s$-wave pairing in which case all quantities are functions of
$k$, the magnitude of $\vec k$, one gets a BCS-like
set of self-consistent equations \cite{Nozieres-Comte,BCSBook}:
\begin{eqnarray}
\xi_k &=& \epsilon_k - \mu - 2 \sum_{ k'} V^{ee}_{ k -  k'}
n_{k'} 
= \epsilon_k - \mu - \sum_{\vec k'} V^{ee}_{ k -  k'}
(1 - \xi_{k'} / E_{k'}), 
\label{SCF1} \\
\Delta_k &=& 2\sum_{ k'} V^{eh}_{ k -  k'} <a_{c, k}^{\dagger}a_{v, k}>
 = \sum_{ k'} V^{eh}_{ k -  k'}
\Delta_{k'} / E_{k'},
\label{SCF2} \\
E_k^2 &=& \xi^2_k + \Delta^2_k.
\label{SCF3} 
\end{eqnarray}

Here Eq. (\ref{SCF1})  gives the renormalized
single-particle energy (per pair) $\xi_k$ measured from the chemical
potential.($\epsilon_k = \frac{k^2}{2m_e} +\frac{k^2}{2m_h}$.)
Eq. (\ref{SCF2}) is the ``gap equation'', familiar from BCS, so
$\Delta_k$ is the gap-function and is also the order-parameter. Note that in order for
$\Delta$ to exist {\em both} $u$ and $v$ must be non-zero for some overlapping range of momenta $k$; this function describes the overall degree of phase-coherence.
$E_k$ can be identified as the pair-breaking excitation spectrum:
it is the energy cost of taking one pair out of the condensate and
placing them in plane-wave states of momentum $\vec k$.

The BCS ansatz is exactly equivalent to a Hartree-Fock approximation,
allowing for the possible (self-consistent) expectation value of an
{\em off-diagonal} self-energy term. 
%
The spectrum of Eq. (\ref{SCF3}) can be seen as arising from the action
\beq
\label{eq:HFA}
\left( \begin{array}{c c} a_{c, k}^{\dagger} & a_{v, k}^{\dagger} \end{array} \right )
\left( \begin{array}{c c} \omega - \frac{1}{2} \xi_k  & \frac{1}{2}\Delta^*_k \\
\frac{1}{2} \Delta_k & \omega + \frac{1}{2} \xi_k \end{array} \right )
\left( \begin{array}{c} a_{c, k} \\ a_{v, k} \end{array} \right)
\enq

If the density is low, $r_s \gg 1$, then the isolated excitons are expected to overlap very little. Hence we expect that $v_k \ll 1$, and $u_k \approx 1$ so that the wavefunction has the approximate form
\beq
\label{eq:wfndilute}
|\Psi_{MF}> \stackrel{r_s \to \infty }{\rightarrow} \prod_{\vec k}\frac{1 + \lambda \phi (k,0) a_{c,k}^\dagger a_{v,k}}
{\sqrt{1+\lambda^2 \phi(k,0)^2}}\left| 0 \right>
\enq
where $\lambda \propto n^{1/2} \propto r_s^{-1}$ is now small. In this
limit $\mu < 0$ (we measure energies from the bottom of the combined
electron and hole bands) and approaches -1 Rydberg as the density
becomes infinitesimal -- just the binding energy of the electron-hole
pair. The lowest excitation energy of the system occurs at $k=0$, and
corresponds to the ionisation of an exciton into a free electron-hole
pair.

In the opposite limit of high density where the electron and hole kinetic energy dominate the interaction energy, we should expect to find a ground state consisting of two interpenetrating Fermi liquids, i.e.
\beq
\label{eq:wfndense}
|\Psi_{MF}> \stackrel{r_s \to 0 }{\rightarrow} \prod_{|{\bk}|<k_F} a_{c,k}^\dagger a_{v,k} \left| 0 \right>.
\enq
So for $r_s \ll 1$ we expect that $v_k = \Theta (|k|-k_F)$, where
$k_F$ is the Fermi momentum of the occupied electrons (or holes). So
here $\mu = \epsilon_{k_F}$ and is positive -- within the bands. In the
extreme limit $r_s \to 0$ the order parameter vanishes; for small,
non-zero $r_s$, the model can be explicated in terms of a Fermi
surface instability. Here, the effect of the Coulomb interaction is
confined only to states close to the Fermi surface, producing a small
rounding of the occupation functions away from those of the free Fermi
gas. The order parameter $\Delta_k$ is small (in comparison to $\mu$)
and generated mostly by states whose momenta are within
$\Delta_{k_F}/v_F$ of the Fermi wavevector, $v_F=\partial \epsilon /
\partial k$ being the Fermi velocity. The
minimum excitation energy equals $\Delta_{k_F}$, and involves
breaking pairs whose components have momenta near to the Fermi surface.

\begin{figure}[htpb]
\begin{center}
\includegraphics[width=12cm]{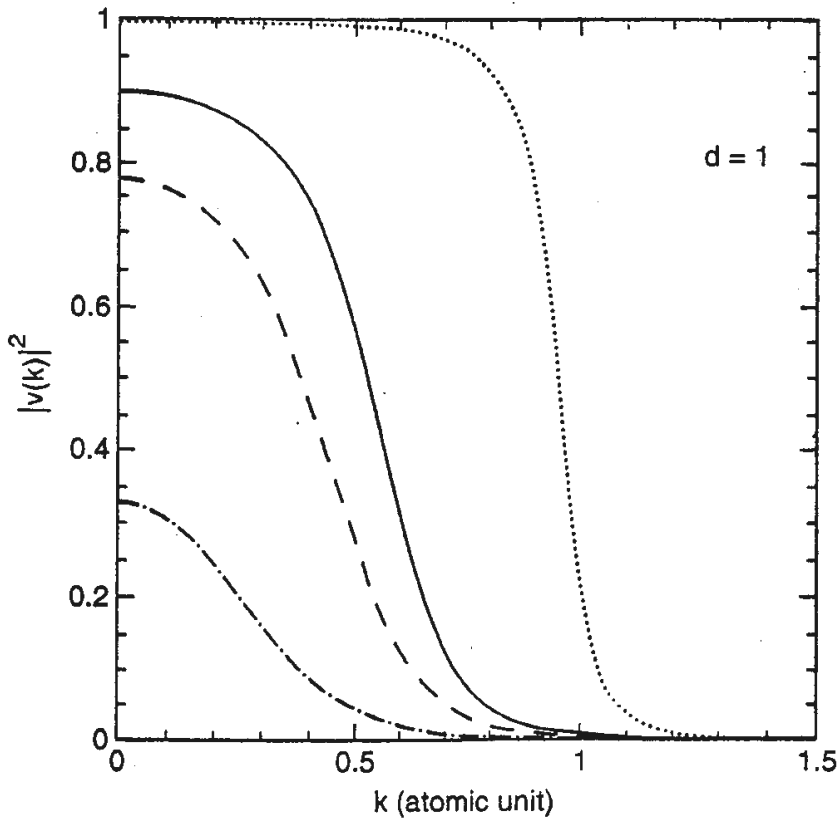}
\end{center}
\caption{\label{fig:vk} Occupancy $v(k)^2$ of the variational wavefunction at low and high densities. Note how it evolves from nearly a Fermi function at small $r_s$ to a Lorentzian form (expected for a hydrogenic exciton) at low density. 
Dotted line: $r_s = 2.11$;
thick solid line: $r_s = 3.69$;
dashed line: $r_s = 4.72$;
dotted-dashed line: $r_s = 9.56$. The calculations are for coupled quantum wells at a separation of 1 Bohr radius. From \protect\cite{PBL96}. 
}
\end{figure}

As an example of how this works in practice, \figref{fig:vk} shows the
evolution of the variational wavefunction from low to high density,
calculated for a bilayer electron-hole system \cite{PBL96}. The trends
we have described above are quite clear, so this ground state
wavefunction apparently does a good job with the oft-called BCS to BEC
crossover, with, however, a wavefunction that is always of the same
form.

\begin{figure}[htpb]
\begin{center}
\includegraphics[width=12cm]{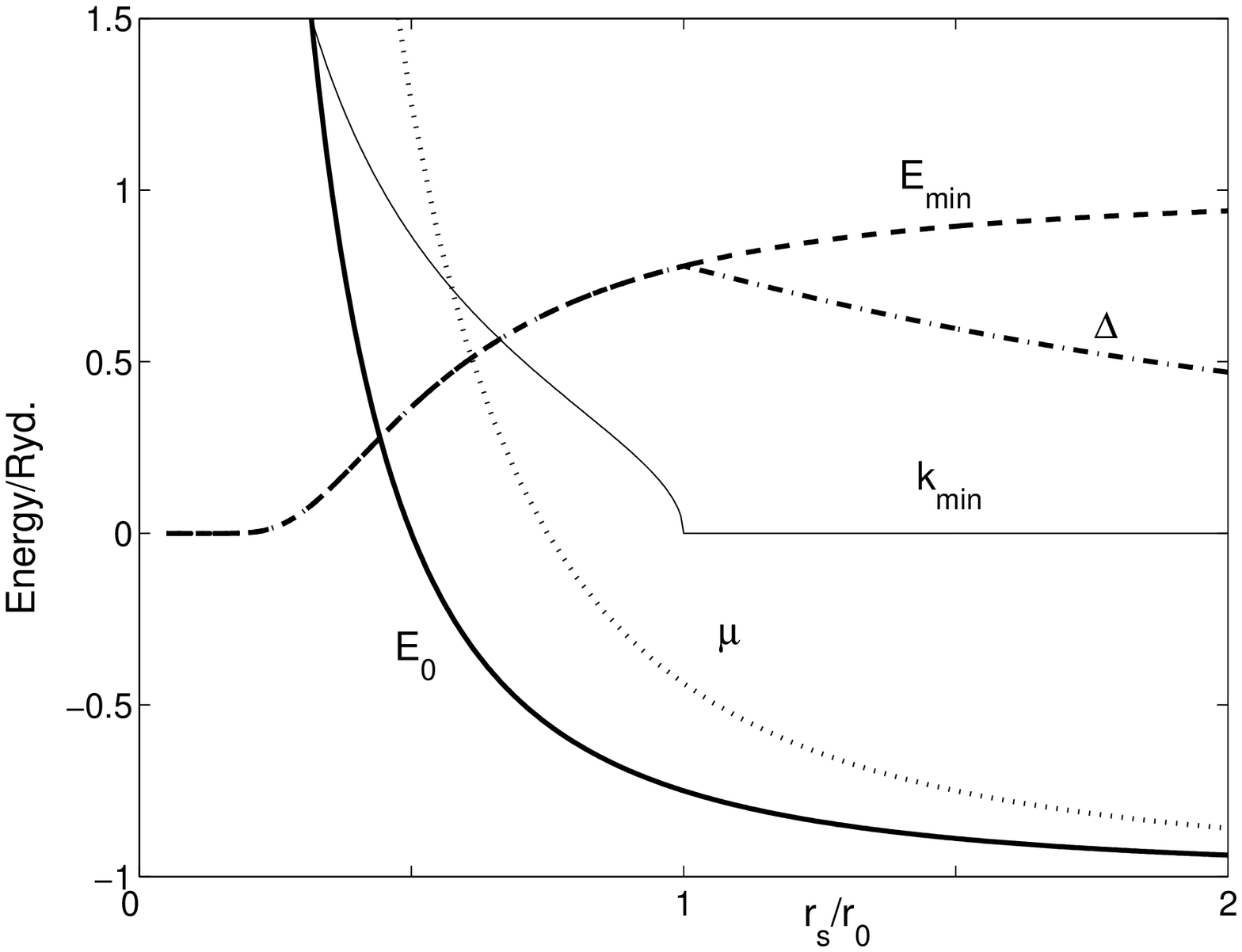}
\end{center}
\caption{\label{fig:energies} Sketch of the energy per particle $E_0$
(solid line) and chemical potential $\mu$ (dotted) relative to the
band edge, along with the correlation gap $\Delta = \Delta(k_{min})$
(dash-dot). Also shown is the minimum excitation energy $E_{min} =
\min(E_k)$ (dashed) and the wavevector $k_{min}$ (think solid line) of
the minimum gap. Typically, the density parameter $r_{0}$ marking the
BCS-BEC crossover is around 2-3.}
\end{figure}

\begin{figure}[htpb]
\begin{center}
\includegraphics[width=12cm]{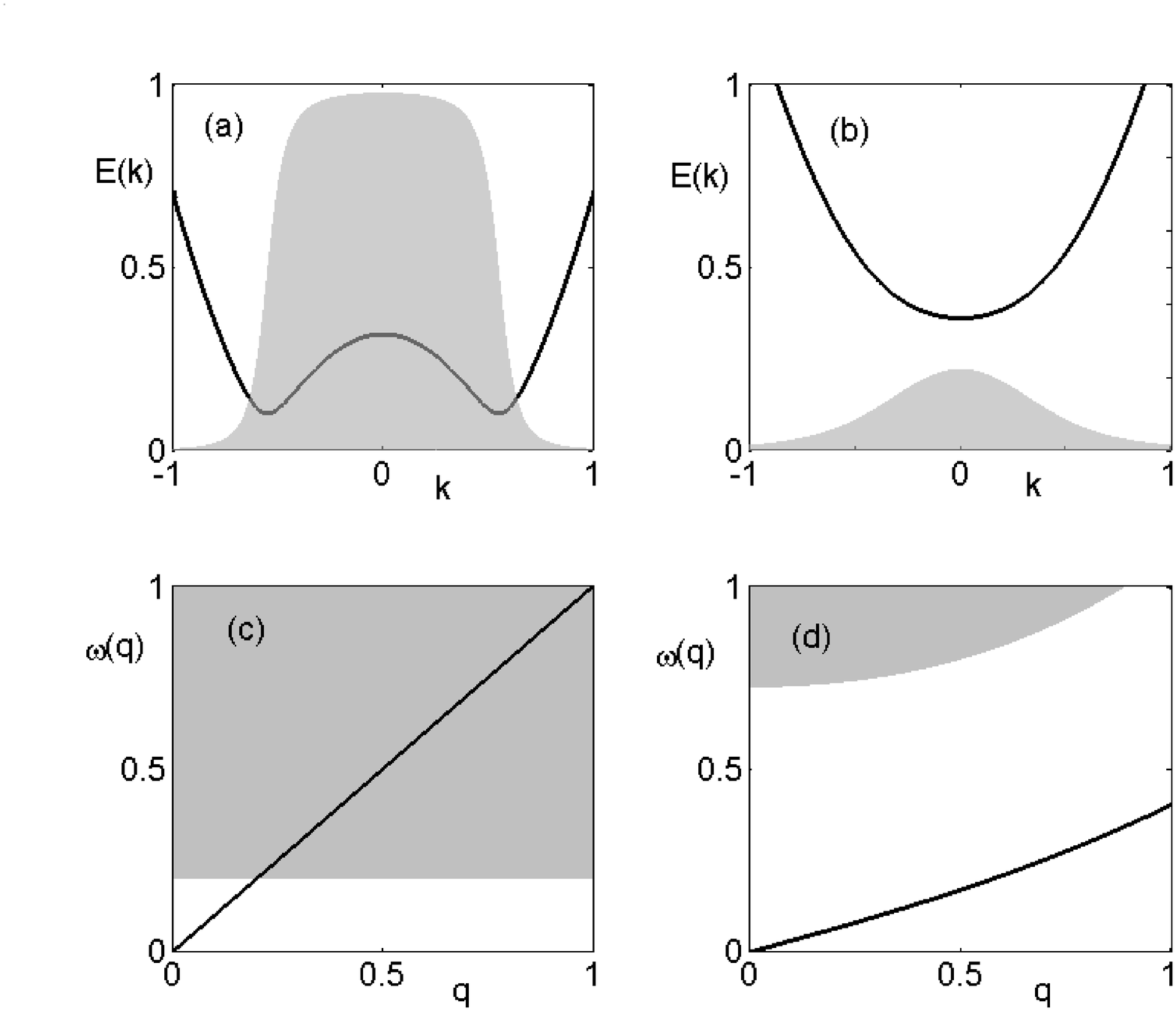}
\end{center}
\caption{\label{fig:excitations} (a) and (b) show a sketch of the
quasiparticle spectrum $E(k)$ (lines) and the occupation factor
$v(k)^2$ (grey hatching) on either side of the BCS crossover. In the
lower panels, (c) and (d) give for comparable regimes the spectrum of
excitations of total momentum $q$.  In the dense (BCS) limit (c) shows
a steeply rising phase mode and most of the phase space is occupied by
gapped particle-hole excitations. In the dilute limit (d) the particle
hole spectrum is at the ionisation energy, and the phase mode provides
the dominant fluctuations.}
\end{figure}

Along with the change in wavefunction, the energy spectrum changes
also. In Fig. \ref{fig:energies} we show a qualitative sketch of the
behaviour of the parameters of the theory as a function of $r_s$. (A
particular calculation for 2D bilayer systems is given in
\cite{PBL96}, which confirms the trends shown here, though details may
differ -- in particular $E_{min}$ may have a weak maximum near the
point where the chemical potential passes through the bottom of the
band.)  As $r_s$ increases see that the chemical potential ($\propto
1/r_s^2$ in the plasma) drops below the bottom of the free
electron-hole band, reaching eventually $-1$ Rydberg as $r_s \to
\infty$. The ground state energy per particle tends also to the same
value, as we expect. Near to where $\mu$ crosses the band edge the
quasiparticle-hole excitation spectrum changes its form, as the
minimum excitation energies go from being near to the Fermi wavevector
$k_F$ to being at $k=0$. The latter excitations correspond just to the
unbinding of an exciton into free particles and holes (see Figure
\ref{fig:excitations}). In the high density limit, the gap parameter
$\Delta(k_{min})$ and $E_{min}$ are the same. In the low density
limit, $\Delta$ becomes small, but $E_{min}$ stays large.

\begin{figure}[htpb]
\begin{center}
\includegraphics[width=12cm]{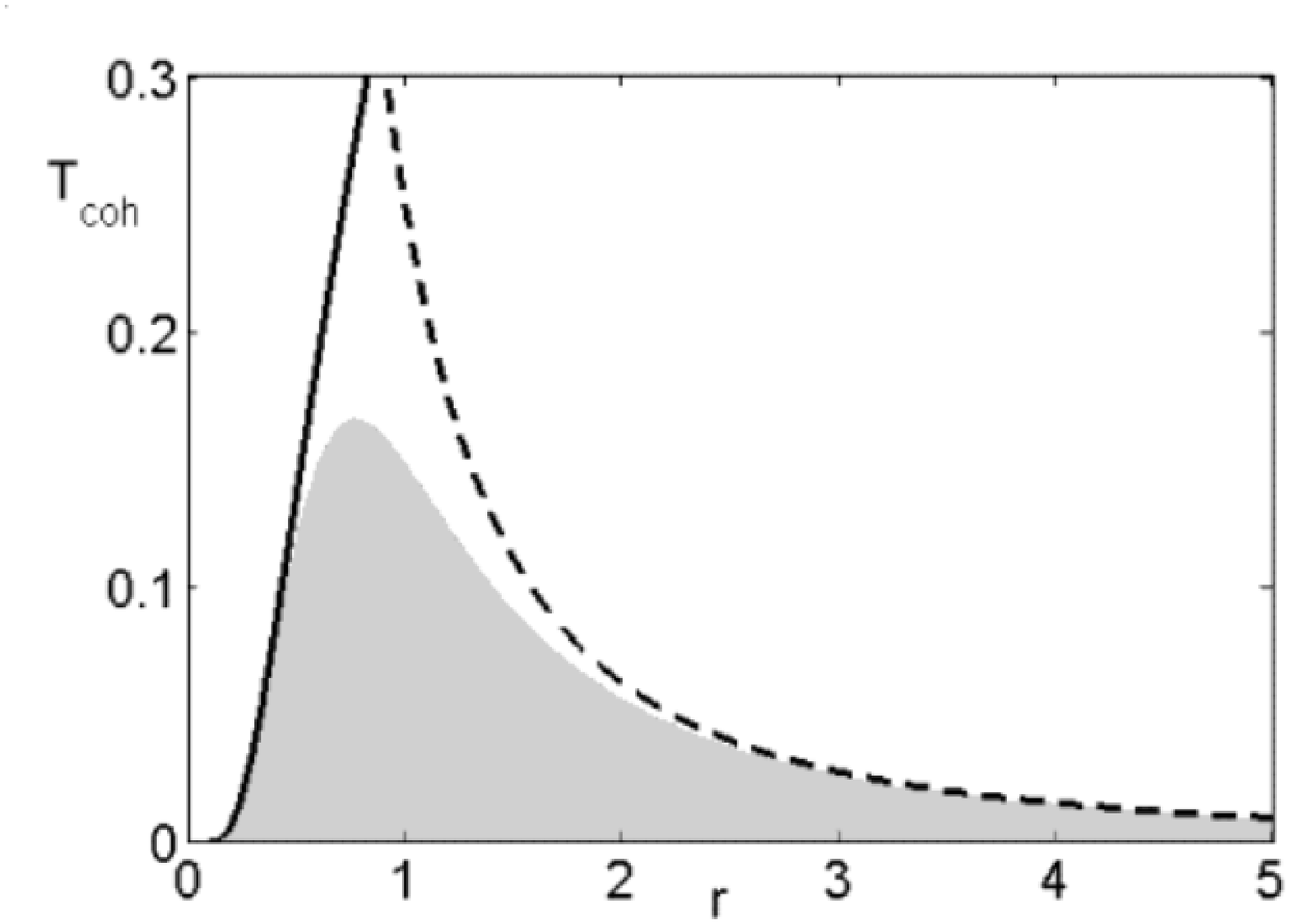}
\end{center}

\caption{\label{fig:bilayertc} Estimates of the coherence temperatures
in $Ryd.$ for the BCS limit (solid line, Eq. (\ref{eq:tcbcs})) and the BEC
limit (dashed line, Eq. (\ref{eq:tcbec})). The parameter $A=1$ and the
scale for the BCS limit has been fit to the calculations of coupled
quantum wells of \cite{PBL96}, and $m_e = m_h$. The gray hatching is a
smooth interpolation between the limits. }
\end{figure}

Although this seems like a sensible treatment of the ground state
wavefunction and low temperature properties, it is a poor theory for
finite temperature. Clearly a BCS theory of $T_c$ will estimate the
transition temperature to be of order $E_{min}$, which is sensible at
large density, but clearly nonsense in the bosonic limit. The error is
well-known -- the BCS excitation spectrum is missing the collective
excitations.  Notice that the excitations in the BCS state are all
pair-breaking excitations with total momentum $q=0$. There is no sign
of the sound mode expected from the Bogoliubov spectrum
(\ref{eq:bogoliubov}), which one would certainly expect to recover in
the dilute limit. This is in fact a traditional problem with
mean-field theories of correlated ground states: for example the
Slater (or Hartree-Fock) theory of magnetism is missing a spin-wave
spectrum; the BCS theory of superconductivity misses the Bogoliubov
phase mode; the mean-field theory of charge- and spin-density waves is
lacking a ``phason'' or sliding mode. In the present problem, notice
that in the low density limit we are also apparently missing all of
the bound exciton excited states, despite that the ground state
wavefunction is of course exact as $r_s \to \infty$.

Conveniently,the problem is also straightforwardly rectified,
following methods that were first developed for superconductivity
\cite{BCSBook}. One method to do this -- that preserves the high energy
structure (on scales of order the gap) as well as giving the
appropriate low energy theory -- is to go back to the complete
derivation of the (single) exciton spectrum (including its centre of
mass motion) by calculating the repeated interaction of an electron
and a hole.
This is discussed carefully by Mahan\cite{Mahan}. For a single exciton
it gives the usual spectrum, with both the free motion of the center
of mass and the series of bound excited states of higher internal
quantum numbers.

%

In the condensed state, one should repeat this calculation but now
using the {\em quasiparticle} propagators of Eq. (\ref{eq:HFA}).  Now we
find that the 1S exciton dispersion becomes {\em linear} at small $q$,
which is the Bogoliubov mode we expected in analogy to
Eq. (\ref{eq:bogoliubov}).  Detailed results have been given by Keldysh
and Koslov\cite{Keldysh-Koslov}, and others \cite{Cote,Bauer}. There
is an equivalent functional field theory approach to this scheme,
which by explicitly preserving the gauge symmetry of the low energy
theory guarantees the correct form of the phase mode
\cite{Marchetti04} and runs close to the line of Section
\ref{sec:collective}.

The algebra can become messy, but the physics in two limits is clear,
and most of the useful results can just be sketched by hand.
\figref{fig:excitations} contrasts the excitation spectrum in the low
and high density limit. At high densities, the phase mode has a steep
velocity $s$ of order the Fermi velocity $\approx v_{f}$, because the
energy of a pair excitation is almost entirely the (large) kinetic
energy of two fermions shifted from the Fermi surface by momentum $q$,
i.e. $s \approx q d \epsilon_k / d k |_{k_f}$. The mode then runs into
the continuum at a momentum of order $1/\xi_{BCS}$ with $\xi = \hbar
v_f/\Delta$ the familiar BCS coherence length. Provided $n \xi^d \gg
1$, the phase space where the sound mode is of lowest energy is small,
and consequently the dominant thermal excitations that destroy the
superfluid order are broken pairs.  In contrast, at low density, the
particle hole gap is large, of order the Rydberg, while the sound
velocity is approximately given by the Bogoliubov result discussed
above \beq M s^2 \approx n_oV_o = 3 \frac{m}{M} \frac{a}{a_o}
\frac{1}{r_s^3} {\rm Ry.}  \enq in three dimensions. Here we have
re-expressed the interaction potential between dilute excitons in
terms of the scattering length via $V_o = 4 \pi \hbar^2 a / M$, $m$ is
the exciton reduced mass, and $M$ the exciton mass. (On physical
grounds one expects $a \propto a_o$, though for long range dipole
interactions between 2D excitons, this approximation may not be used.)
The linear dispersion turns quadratic for momenta larger than the
inverse of the healing length, which is \beq \frac{\xi}{a_o} = \left (
\frac{r_s^3 a_o}{6a} \right )^{1/2}. \enq So in this limit, the phase
mode turns smoothly into the kinetic energy of the 1S exciton; it
never intersects the continuum, instead running parallel to it. (There
also exists the Rydberg series of excited states of the pairs,
neglected here for simplicity.)

We can now estimate the crossover in the transition temperature from
dense to dilute limits, expressed in exciton Rydbergs for
convenience. In the BCS limit we will get \beq
\label{eq:tcbcs}
\frac{kT_c}{Ry.} \approx e^{-1/g} \approx e^{-A/r_s} \;\;\; {\rm for }
\; r_s \ll 1, \enq where $g \approx V_{eh}(k_f)/E_f \propto r_s$
and $A$ is a constant of order unity. In the dilute limit we shall
have a transition temperature of order the degeneracy temperature in
the non-interacting Bose gas \beq
\label{eq:tcbec}
\frac{kT_c}{Ry.} \approx \frac{m}{M} \frac{1}{r_s^2} \;\;\; {\rm for }
\; r_s \gg 1. \enq Thus $T_c$ is a strong function of density
peaking near $r_s \approx 1$, and vanishing in both low and high
density limits.

An estimate for bilayers is shown in \figref{fig:bilayertc}. Since the
system is two-dimensional the actual transition will be of
Kosterlitz-Thouless character, and thus reduced by a numerical factor
from the mean-field estimates given here. More important than the
quantitative changes in $T_c$ is here the fact that long-range order
will not occur at any non-zero temperature, because although there is
the rigidity provided by the acoustic mode, thermal fluctuations of
the phase mode decorrelate the phase of the order parameter. This has
pronounced effects on the phase-coherent emission of
light\cite{Keeling03}.

\subsection{Miscellaneous remarks}

We make a few small remarks and caveats about the solutions here.

Because we used a bandstructure model with isotropic dispersion, the
electron and hole Fermi surfaces are always perfectly nested, and
therefore even at infinite density there is a nesting instability of
the Fermi seas to an excitonic insulator with a tiny gap. This is
suppressed by realistic bandstructure effects -- for example in GaAs
the hole bands are anisotropic, being based on p-orbitals -- so that
there is a sharp onset of $T_c$ at a critical density. Once the
Coulomb interaction is itself a sizeable fraction of the kinetic
energy, the transition is no longer driven by a nesting instability.

The BCS wavefunction itself gives a poor bound for the overall energy
of the ground state, largely because it neglects the short range
correlation of like species. Improved wavefunctions of the Jastrow
form\cite{Zhu96,Senatore} give lower energies without destroying the
qualitative description encapsulated by the BCS state. In particular,
there appears to be no stable electron-hole liquid state in a
direct-gap semiconductor (i.e. a minimum in the ground state energy
per particle at large density, below the binding energy of exciton or
biexciton), unlike the case of the indirect gap Ge\cite{ehliquid}.

Bilayers are particularly advantageous in that the dipole repulsion
between individual excitons strongly disfavours biexciton formation.
In order to prepare a quasi-equilibrium state of excitons not under
direct illumination, it is necessary to prepare traps, perhaps by
ambient disorder\cite{Butov94,Timofeev}, well-width
fluctuations\cite{Zhu95}, or strain\cite{Snoke99}. These all turn out
to be relatively shallow, and the density distribution of excitons
changes very little through the condensation transition
\cite{Keeling03}. Thus, in contrast with the cold atom systems, the
direct spatial imaging of density is not expected to provide dramatic
evidence for condensation.

We have ignored spin, and of course excitons made of s=1/2 fermions
will come in singlet ($L=0$) and triplet ($L=1$) varieties. In GaAs
and similar systems, because the (spin-orbit coupled) heavy and light
hole states have $J=3/2$, there are optically active excitons
with angular momentum $L=\pm 1$ as well as dark excitons with $L=\pm
2$. In quantum wells, the broken degeneracy between heavy and light
hole bands yields two energetically well-separated exciton
species\cite{Chuang}. In the bilayer quantum well systems, interband
exchange is certainly much too small to give significant energetic
splitting between spin species, thus if equilibrium is established
between the spin species the only effect is to replace $r_s \to
g^{1/2} r_s$, with $g$ the spin degeneracy\cite{Nozieres-Comte,Zhu95}.

We stress again the neglect of tunnelling and recombination. There are
systems of type II heterostructures (e.g. InAs/GaSb) where the
conduction band of one material lies below the valence band of the
other. Thus an interface between the two will produce a pair of
inversion layers (electrons and holes) in close proximity. Generally,
the overlap between electron and hole will not be negligible, so that
tunnelling terms $t c^\dagger v$ will exist in the Hamiltonian, and
exciton conservation is destroyed. Firstly, this will introduce a gap
in the spectrum even without Coulomb correlation (the system may
become an insulator or semimetal)\cite{Lakrimi}. More generally, the
gauge symmetry is broken so that the order parameter $<c^\dagger v>$
has its phase fixed by the tunnelling matrix element, and the
Bogoliubov mode has a gap.  Only should the tunnelling be vanishingly
small (as it may be in the quantum Hall bilayer systems
\cite{EisensteinBilayer,MacDonaldBilayer}) can one expect to approach
superfluid behaviour.

\section{Theory of polariton condensation}
Excitons are of course excitations above the ground state -- so in
order to work with an out-of-equilibrium ensemble in the previous
section we introduced a chemical potential and enforced thermal
equlibrium.  But in many semiconductors, there is a direct
recombination channel of excitons into dipole radiation, which is
suppressed but not eliminated, for example, in the bilayer systems,
because recombination requires tunnelling between the coupled quantum
wells.

The decay of excitons into photons can of course provide evidence for
the coherence in the exciton system, both
temporal\cite{Butov94,Dicke-ferroelectric} and
spatial\cite{Keeling03}. If the coupling is weak, as in the coupled
quantum wells, or in $\mathrm{Cu_2O}$, then the exciton system is only lightly
perturbed by the decay process. However, there is a different limit of
strong coupling that can be obtained by exciting excitons inside
optical microcavities\cite{Microcavity}. If the photons are
well-confined by mirrors, then the appropriate linear excitation is a
superposition of photon and exciton, called a
polariton\cite{Hopfield}. This is a new type of boson, and on account
of its light mass, seems a natural candidate for polaritonic
BEC\cite{Hanamura-Haug,Kavokin02,Kavokin03} at substantial
temperatures. Of course, since photons are not conserved, we must
again consider the quasi-equilibrium situation of a pumped system with
(nearly perfect) mirrors that has attained thermal equilibrium with a
bath that establishes a chemical potential for the excitation number.

Free photons in the cavity are described by the microscopic quasi
two-dimensional Hamiltonian
\begin{equation}
  H_{\mathrm{ph}} = \sum_{\bp}
  \psi_{\bp}^\dag\left[\omega (\bp) - \mu\right]
  \psi_{\bp} \; ,
\label{eq:photo}
\end{equation}
where their dispersion, $\omega (\bp ) = \sqrt{\omega_c^2 + (c \bp
)^2}$, is quantised in the direction perpendicular to the plane of the
cavity mirrors, and we shall just keep a single branch of the cavity
modes, beginning at $\omega_c = c \pi /L$ (whose value is fixed by the
cavity thickness $L$).

In the dipole and rotating-wave approximation, the photons are assumed
to be coupled to the electron-hole system through a local interaction,
\begin{equation}
  H_{\mathrm{dip}} = g \int d{{\bf r}}\left[\psi
  ({\bf r} ) a_c^\dag ({\bf r} ) a_v({\bf r}) + \mathrm{h.c.}\right].
\end{equation}
In practice, one chooses $\omega_c$ to be close to the exciton
frequency so the resonant coupling dominates.  Since we are dealing
with a system where the physical temperature is much smaller than the
photon frequency $\omega_c$, we may neglect the tiny spontaneous
population that would be generated by non-resonant terms.  To mimic
the effect of the external excitation source, we suppose that the
electron-hole/photon system is held in quasi-equilibrium by tuning the
chemical potential $\mu$ in~Eq. (\ref{eq:chemical}) to fix the total
number of excitations
\begin{equation}
  \hat{N}_{\mathrm{ex}} = \sum_{\bp} \psi^\dag_{\bp}
  \psi_{\bp} + \frac{1}{2} \sum_{\bk}
  \left(a_{c,k}^\dag a_{c,k} -
  a^\dag_{v,k} a_{v,k}+1\right)\; .
\label{eq:nexci}
\end{equation}
However, how the system chooses to portion the excitations between the
electron-hole and photon degrees of freedom depends sensitively on the
properties of the condensate.

In the previous sections, we were at pains to stress the difference
between the statistical physics of BEC of non-interacting bosons, and
the phase transition accompanying coherence. A single polariton is a
phase-coherent object, delocalised over the whole system and producing
a coupled oscillation in the electric displacement field ${\bf D}$ (of
light) and the excitonic polarisation ${\bf P}$. Polariton
condensation would lead to a {\em macroscopic} coherent optical field
in the cavity (phase-locking of the polariton modes), and hence bear
considerable similarity to a
laser\cite{Hanamura-Haug,Dicke-ferroelectric}. What is special about
the condensed polariton state is that the excitonic component is also
coherent, whereas this is strongly dephased in a conventional laser,
and only a coherent photon field exists.

For strongly detuned excitons and photons, exciton-photon condensation
can be described either in terms of polariton condensation or as
exciton condensation with both the Coulomb interaction and a
photon-mediated interaction. If the excitons are localised, we expect
the photon-mediated interaction to dominate, because its range is
usually larger than that of the Coulomb interaction between excitons.

\subsection{Mean-field wavefunction}

There is now a very natural extension of the Keldysh mean field
wavefunction to propose for the coupled problem, viz.
     \begin{equation}
     \label{varwfn2}
     |\Psi_0> = e^{\lambda \psi_0^\dagger}\prod_{\vec k} [ u_{\vec k} + v_{\vec k} a_{c, k}^{\dagger}
                                                   a_{v, k} ]|0\rangle.
     \end{equation}
     Now one has, in addition to the variational functions $u,v$, a variational parameter $\lambda$. This is a state which is
     a coherent state of photons (in the lowest mode of the cavity), and a coherent state of excitons. The equations which arise
     from a variational minimisation of $\langle \Psi_0|H|\Psi_0\rangle $ couple these order parameters, and the relative proportions
     of photon and exciton in the ground state depend on details such as the relative tuning of the exciton and photon energy; but
     both take macroscopic values in the state $|\Psi_0\rangle$ of Eq. (\ref{varwfn2}). 

        The variational equations can be found elsewhere\cite{Marchetti04}, and we will here just discuss the results qualitatively.
Just as the Keldysh wavefunction, Eq.\ (\ref{eq:keldyshwfn}),
     approximates a condensation of structureless excitons in the low-density limit ($v \ll 1$), in the same limit
     Eq.\ (\ref{varwfn2}) will look like a Bose condensate of polaritons. In the dense limit, $v_k$ approximates a Fermi function and
     only close to the chemical potential is there any renormalisation of the spectrum. If one detunes the photon frequency far
     from the chemical potential (i.e. $|\omega_c - \mu| \gg g \lambda$) the results are barely changed from the old mean field theory because the interaction is dominated by direct Coulomb forces;
     but in
     the opposite limit,
     \begin{equation}
     \frac{g^2}{|\omega_c - \mu|} \gg Ry^*,
     \end{equation}
     the Coulomb interaction is not the relevant source of pairing,
     instead it is the photon field itself.

As far as the electronic excitations which form the condensate are
concerned, they are then identical to those predicted by the
well-known Hartree-Fock theory of a semiconductor in an {\em external}
classical time-dependent field\cite{Galitskii,ACStark}.  The most
obvious difference from the driven problem is just that the photon
field has to be established self-consistently, but this is just a
(complex) technical matter.  A more hidden (and more important
difference for the robustness of the state) is that the excitation
spectrum for the quasi-electron and quasi-hole is occupied according
to equilibrium (fermionic) statistics.

\subsection{Localised exciton model}
A simplified model that replaces the excitons by localised two-level
systems is a good way to exhibit the physics in the photon dominated
regime.

The model is the Dicke model of atomic physics\cite{DickeModel}:
     \begin{equation}
     \label{H_2level}
      H_{2level} = \sum_{\bf q} \omega (\bq) \psi_{\bq}^\dagger \psi_{\bq} + \sum_{j=1}^N \frac{\epsilon_j}{2}(b^\dagger_jb_j-a^\dagger_ja_j) +
          \frac{g}{\sqrt{N}}\sum_{j \bq }(b^\dagger_ja_j\psi_{\bq} + \psi^\dagger_{\bq}
          a^\dagger_jb_j)\;\; .
     \end{equation}
     $H_{2level}$ describes an ensemble of N two-level oscillators with an energy $\epsilon_j$ dipole coupled to one cavity
          mode.  $b$ and $a$ are fermionic annihilation operators for an
          electron in an upper and lower states respectively
          (with a local constraint $b_j^\dagger b_j + a_j^\dagger a_j = 1$ so that there is an electron either in the lower level or
          in the upper level) and
          $\psi$ is a
          photon bosonic annihilation operator. The operator that counts the number of excitations in the system,
          $N_{ex}= \sum_q \psi_{\bq}^\dagger\psi_{\bq} + \frac{1}{2}
          \sum_j (b^\dagger_jb_j-a^\dagger_ja_j+1)$, 
commutes with $H_{2level}$ so is conserved.

        The mean field wavefunction is then
        \begin{equation}
          \label{varwfn3}
          |\lambda,u,v\rangle = e^{\lambda \psi_0^\dagger }
               \prod_{j} ( v_{j} b_j^{\dagger} + u_{j} a_j^{\dagger} )
               |0\rangle.
          \end{equation}
        with the (real) variational parameter $\lambda$ and variational functions $v_j= v(\epsilon_j )$. (The vacuum state is here defined to be empty of both levels.) The constraint is satified
by setting $u_j^2+v_j^2 = 1$, and the variational functions are obtained by minimising $H_{2level}-\mu N_{ex}$. For detailed results see \cite{Eastham00,Eastham01}. Notice that this approximation neglects coupling to all but the ${\bq}=0$ photon mode at $\omega_c$.

        To connect to the earlier theory of pure fermions, consider the case when $\delta=(\omega_c -\epsilon)/g \gg 1$. Now provided the occupation is fairly small (less than or order of 1 per site), the chemical potential will lie in the band of two level systems, the photon occupation will be small, and the photons will act to provide a virtual interaction between the excitons of magnitude $g_{eff} = g^2/(\omega_c - \mu)$.

\begin{figure}[htpb]
\begin{center}
\includegraphics[width=12cm]{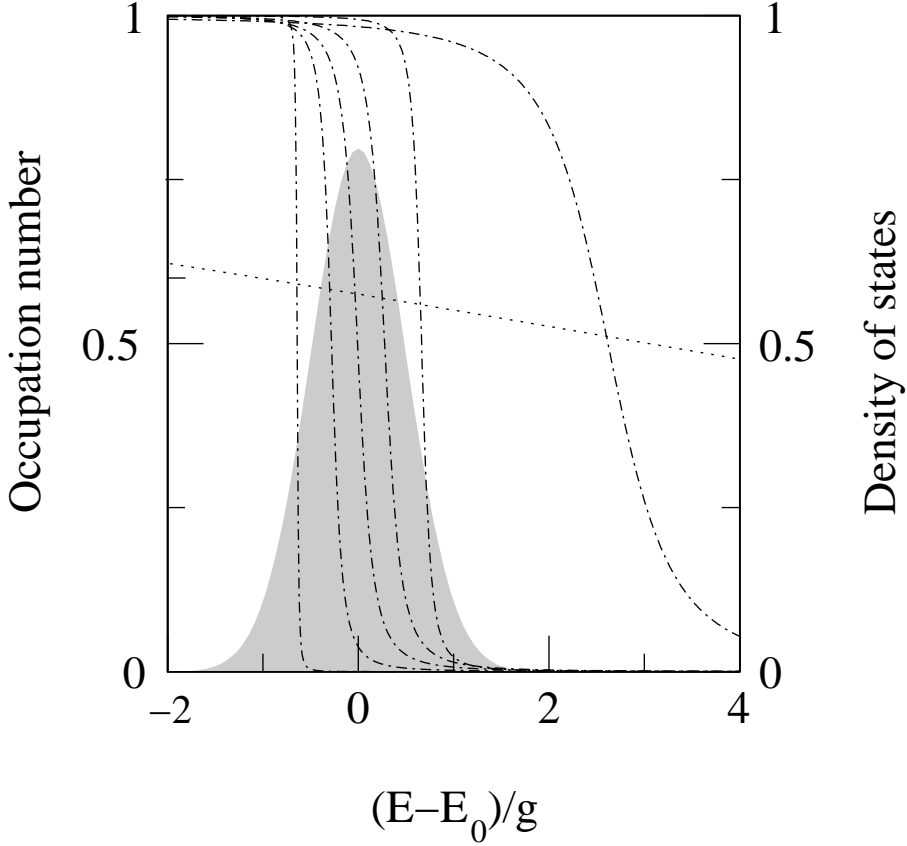}
\end{center}
\caption{\label{fg:vj} Occupancy $v(\epsilon )$ as a function of two level system energy $\epsilon$ where the photon energy is
substantially detuned (above) the centroid of the exciton distribution ($\omega_c - <\epsilon> = 3g$). The different curves correspond to
$\rho_x = 0.1, 0.3, 0.5, 0.7, 0.9$ (dot-dash increasing from left to right) and $\rho_x=101$ (dotted curve). The grey hatching is the density of states $\nu (\epsilon)$ of the two-level systems. From \protect\cite{Eastham01}.}
\end{figure}

        The results are most easily visualised with a distribution of energies, and in \figref{fg:vj} are shown the occupancies calculated for a gaussian distribution of energy levels, as the excitation level $\rho_x=N_{ex}/N$ is increased. Notice that at low densities, the distribution approaches the step function of a Fermi distribution, and becomes broadened as the density {\em increases}, counter to the results of the Coulomb problem in \figref{fig:vk}. The reason is that the gap in the two-level model is not fixed but is provided by the photon field, whose amplitude is growing with $\rho_x$; for $\rho_x > 1$ the order parameter becomes increasingly photon-like. In fact as $\rho_x \to \infty$, then $v^2 \to \frac{1}{2}$ --- the system saturates with the two-level distribution held just above the border of inversion. When the photon and exciton are detuned from each other (as in the case shown in the figure) this evolution is not monotonic, because the chemical potential jumps discontinuously from being within the band of two level systems to be close to the photon.

Just as in the exciton case, we can extend the mean-field theory to
finite temperatures by solving the self-consistent equations assuming
a thermal occupancy of quasiparticles, as in BCS theory. The
transition temperature is determined by setting $\lambda=0$ in the
BCS-like gap equation
\begin{eqnarray}\label{pol-gapeqn} \frac{1}{g_{eff}}= \int
\frac{\tanh(\frac{\beta E(\epsilon)}{2})}{E(\epsilon)} \nu(\epsilon)
d\epsilon\\ E(\epsilon)=\sqrt{(\epsilon-\mu)^2+4|\lambda|^2},
\end{eqnarray} where $\nu(\epsilon)$ is the density of states of the
two-level oscillators. If $\mu$ lies in the band of two-level
oscillators then at low temperatures(relative to the bandwidth of
these oscillators), the integral on the right of (\ref{pol-gapeqn}) is
approximately $-2\nu(\mu) \ln(\beta \theta)$, where $\theta$ is a
cut-off associated with the bandwidth of the oscillators. This gives
us an approximate expression for the transition temperature \beq
\label{tmfpolariton}
\ln {\frac{kT_{mf}}{\theta}} \approx \frac{-1}{2 g_{eff}\nu(\mu)},
\enq valid when the computed $T_{mf}$ is small compared with the
bandwidth $\theta$.  If instead the temperature is large compared with
the bandwidth, we can express the transition temperature in terms of
the dimensionless detuning $\delta=(\omega_c-\epsilon)/g$, the density $\rho_x$, and the
coupling g as \beq
\label{tmfpolariton-nowidth} kT_{mf}=g f(\rho_x,\delta)=g \frac{\delta \pm \sqrt{\delta^2-8\rho_{x}+4}}{4 \tanh^{-1}
(2\rho_{x}-1)}. \enq 

The normal state of this model is an incoherent population of
excitons. The phase transition occurs when the chemical potential for
the excitons crosses the energy of a coupled exciton-photon mode of
zero wavevector. At low densities, the energy of the lowest coupled
exciton-photon state is just that of the conventional, linear response
polariton $E_{LPB}$, and the two-level systems are occupied according
to a Boltzmann distribution. Thus the critical density should be
$\rho_x = e^{-\beta(\epsilon-\mu_c)} = e^{-\beta(\epsilon-E_{LPB})}$,
which is indeed the low-density limit of
(\ref{tmfpolariton-nowidth}). At higher densities the form changes,
because the occupation of two-level systems is no longer a Boltzmann
factor, and because the energy of the coupled exciton-photon mode is
renormalised by the occupation of the two-level systems.

There are some unusual features of the phase diagram of this model
that are produced by the saturable nature of the excitons. One of
these is the multivalued phase boundary (\ref{tmfpolariton-nowidth}),
whose two values correspond to the chemical potential crossing either
of the two coupled exciton-photon modes of zero wavevector. One might
expect the higher energy crossing to be irrelevant, as the system
would already have condensed before it is reached. This is not
necessarily true, however, because the exciton entropy decreases with
increasing density when $\rho_x>0.5$. Thus, at high enough
temperatures, the system can be stable against an excitation which
increases the density, even if it decreases the energy. In the region
where the normal-state entropy decreases with increasing density,
$0.5<\rho_{x}<1$, we find that the normal state is stable if its
chemical potential lies above the lower polariton and below the upper
polariton. Another peculiarity is that for $\rho_x>1$ the saturation
forces some of the excitation into the photon, so the system is
condensed at any temperature.

We now discuss the general behaviour of the transition temperature in
the case of a finite bandwidth and a cavity mode lying well above the
band. At low densities the chemical potential will lie towards the
bottom of the band, $g_{eff}$ will be small, and if the band is broad
enough the weak-coupling form (\ref{tmfpolariton}) will apply. As we
increase the density the chemical potential rises, and the transition
temperature increases exponentially as the density of states and
$g_{eff}$ increases. If the band is broad and the detuning large
enough, the weak-coupling form would continue to hold right through
the band. After $\mu$ has moved through the centre of the band the
density of states begins to decrease, and this could produce a
decrease in $T_{mf}$, although it could be offset by the increasing
$g_{eff}$.  As the density is further increased towards $\rho_x=1$,
the weak-coupling form breaks down as $\mu$ moves into the upper tail
of the band. The chemical potential rapidly jumps up to near the
photon frequency, and the transition temperature diverges according to
the strong-coupling form (\ref{tmfpolariton-nowidth}).

The weak-coupling scenario bears some comparison to the {\em
high-density} Coulomb coupled exciton condensate, because in both
cases the effective interaction is small compared with the
bandwidth. The differences arise because $g_{eff}$ increases with
increasing density, and because the density-of-states is a function of
density. Thus while in the Coulomb-coupled condensate the transition
temperature either saturates or decreases with increasing density,
depending on whether we include screening or not, here we find more
complicated behaviour.

Let us describe two other scenarios for $T_{mf}$. Suppose first that
we keep the photon above the band, but reduce the detuning or
bandwidth. Then there will be a region of density where the
weak-coupling form fails, and we need either the strong-coupling form
(\ref{tmfpolariton-nowidth}) or the full solution to the gap equation
(\ref{pol-gapeqn}). Or consider the case when the photon is below the
peak of a broad band. Then as the density increases the transition
temperature simply crosses from the weak- to strong- coupling forms,
diverging as $\mu \to \omega_c$.

While in the Coulomb problem the mean-field theory is only expected to
hold in the weak-coupling limit, we expect the mean-field theory of
polariton condensation in systems with localised excitons to be more
generally valid. This is because the photons provide a long-range
interaction between the excitons, so we expect mean-field theory to be
a good approximation. It is interesting to note that while the
mean-field theory is an approximate theory for the extended system,
for a model which has only a single photon mode (i.e. a
zero-dimensional microcavity), it becomes exact in the thermodynamic
limit ($N \to \infty$; $\rho_x \to {\mathrm const.} >
0$)\cite{Eastham01}. There has been progress on solving that model at
finite $N$\cite{Vadeiko}.

\subsection{BEC to polariton laser to BCS crossovers}
Because we worked with only a single mode of the electromagnetic
field, our discussion of polariton condensation makes no mention of
the polariton effective mass. The theories of polariton condensation
we have discussed have the character of BCS theory, in that finite
temperatures destroy the order by creating excitations across the
gap. In the two-level model this gap, which plays the role of the
superconducting gap $\Delta$ in BCS theory, is $g<\psi>$, whereas in
the electron-hole model the gap will involve both the
optically-mediated interaction $g$ and the Coulomb interaction. Either
way, the transition temperature in these theories is determined by an
interaction strength, and not by an effective mass as it would be were
we to regard the polaritons as structureless, weakly-interacting
bosons. In that theory, we would expect the onset of coherence at a
temperature  \beq
\label{Tbecpolariton}
k_B T_{BEC} \approx \frac{\hbar^2 \rho_{x}}{2 M^*}=\frac{\hbar
c^2\rho_x}{4\omega_c},
\enq
where we have substituted for the polariton mass $M^* = 2 \hbar
\omega_c/c^2$ in the case of resonance, i.e. $\omega_c=\epsilon$. 
This temperature increases rapidly with density since the polariton
has a very light mass: $M^*/m = 2 \hbar \omega_c/(mc^2) \approx
10^{-5}$. But of course it then rapidly reaches a scale of order $g$
when the dominant fluctuations are not the long-wavelength phase
modes, but excitations across the gap. To estimate where the crossover
occurs, we introduce the dimensionless density in the usual fashion
$\pi r_s^2 a_*^2 = 1/\rho_x$ so that we can rewrite
Eq. (\ref{Tbecpolariton}) as
\beq
\frac{k_B T_{BEC}}{g} \approx \frac{Ry^\ast}{g} \frac{m}{M^*} \frac {1}{2 \pi r_s^2}.
\enq
Thus polariton BEC in the conventional sense is expected to be the
appropriate theory only for $r_s > 100/(g/Ry)^{1/2}$; the promising
experimental systems all have coupling constants no more than a few
$Ry.$ and so the regime of applicability is small indeed. At higher
excitation levels the relevant theory is then the mean field theory of
the last section. These estimates for the crossover differ somewhat
from those made in \cite{Kavokin02}.

Of course, as we saw in the last section, once the system reaches
substantial photon densities (approaching the conventional inversion
point for the laser), the mean-field theory gives an unphysical
infinite transition temperature. This implies that there must then be
a second crossover {\em back} to a regime where fluctuations into
states of finite momentum are important. Because lower branch
polaritons at large momentum (outside the light cone) are essentially
excitons uncoupled to the photon bath, this reservoir has a very large
density of states that depletes the condensate and reduces the
transition temperature\cite{Keeling04}.

We now see that there is typically a substantial regime where may find
a polariton condensate in the strong coupling regime but where $r_s
\gg 1$; here our approximation of replacing mobile excitons by
localised two-level systems can be a good one. Yet if $r_s$ is small
enough (or at least that part of the density that is excitonic in
character) we will have to deal with a realistic model of exciton
unbinding - the Coulomb interaction will play a role. This will
produce a second crossover akin to that discussed in
\secref{sec:BCS-BEC}.  Nevertheless, even here there will be a regime
where the photon field will dominate the Coulomb interaction, to be
reached at high excitation levels\cite{Marchetti04}.

\begin{figure}[htpb]
\begin{center}
\includegraphics[width=12cm]{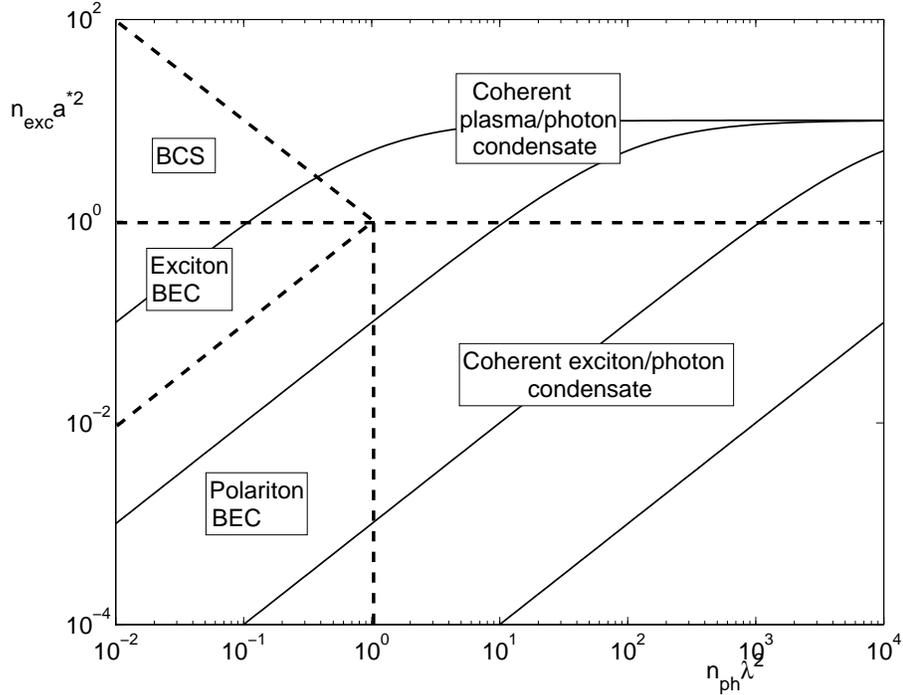}
\end{center}
\caption{\label{fig:phasediag} Sketch to demonstrate the various crossovers in the polariton problem. The dimensionless densities are plotted on the two axes $n_{exc}a^{*2}$ for excitons and $n_{ph} \lambda^2 = n_{ph} \hbar^2/2M^*g$ for photons.
The crossover from BEC of polaritons to an interaction-driven
polariton condensate occurs when $n_{ph} \lambda^2 \approx 1$; the
conventional BEC/BCS crossover for excitons occurs when $n_{exc}a^{*2}
\approx 1$, and at large photon numbers this marks the conventional
point of inversion for a plasma ``laser''. The two wedges labelled BCS
and exciton BEC consitute the regime where the Coulomb interaction is
the dominant coupling term. The solid lines are rough guides to
trajectories that would be followed for a fixed ratio of the coupling
constants $(a^*/\lambda)^2= (g/Ryd.)(M^*/m) =
10,\;10^{-1},\;10^{-3},\; 10^{-5}$. In order for the electron-hole
density to be able to reach such high values as shown, the cavity mode
frequency would need to be placed well above the edge of the band.}
\end{figure}

Figure \ref{fig:phasediag} provides a rough and ready estimate of the
various regimes that may appear for delocalised excitons together with
the coupling to photons in a microcavity. The vertical axis is the
direction of the conventional BCS-BEC crossover of
\secref{sec:BCS-BEC}. However, if there is coupling mediated by
photons, this will always dominate in the model both at very low
density and very high density --- the photon-mediated coupling is
finite and long range, whereas the direct Coulomb coupling is
irrelevant in the two extreme limits. Of course in the physical
system, one cannot tune independently the photon density and the
exciton density, because these adjust their balance to maintain a
common chemical potential.

\subsection{Decoherence and disorder}

Is indeed the polariton condensate just a laser? In fact it differs
very much from the conventional textbook description, which has a
coherent optical field (ignoring finite size fluctuation effects) but
is not generally thought to have a coherent internal polarisation. The
usual assumption of laser physics is that the electronic polarisation
is described by a Langevin equation with a short decoherence
time\cite{laserbook}. There is also a distinction to be made between
most solid state lasers where the electronic excitations are localised
(usually atomic in nature, and thus describable as localised excitons)
and GaAs semiconductor lasers that are usually operated in a regime
where the excitons are unbound (corresponding to a hot two-component
plasma)\cite{sclaserbook,Haug-Koch}. But notice that the latter
distinction is quite independent of whether or not the electronic
polarisation has a short decoherence time -- in principle either a
plasma or an array of two-level systems can support a coherent
polarisation.

The origin of decoherence is elastic and inelastic scattering whereby
the fundamental excitations are coupled to continuum degrees of
freedom in an open system. There are many sources of decoherence:
Because the mirrors are not perfect, light will leak out of the lasing
mode and the excitation (in steady state) must be replaced by
incoherent pumping of excitons; Excitons themselves may decay
spontaneously into photon modes other than the cavity mode; Phonons
and disorder inside the material can scatter the excitons, and produce
pairbreaking and dephasing. All of these may be modelled by coupling
of the internal degrees of freedom to (bosonic) baths of dynamic
fluctuations $B_\gamma ({\bf r} , t )$. If we consider the Dicke model
(\ref{H_2level}), but relax the local single occupancy constraints
$b_j^\dagger b_j + a_j^\dagger a_j =1$, then these fluctuations will be of three
generic types:
\begin{eqnarray}
\label{eq:baths}
      H_{SB} &=&  \sum_{j=1}^N (b^\dagger_jb_j-a^\dagger_ja_j)(B_{1j}^\dagger + B_{1j}) + \nonumber \\
        &+&  \sum_{j=1}^N (b^\dagger_jb_j + a^\dagger_ja_j)(B_{2j}^\dagger + B_{2j}) + \nonumber \\
        &+&  \sum_{j \bq }(b^\dagger_ja_j B_{3j}^\dagger + a^\dagger_jb_j B_{3j}).
\end{eqnarray}
This is already a simplification in that we have kept just diagonal terms. The three terms in Eq. (\ref{eq:baths}) correspond to neutral, pairbreaking, and phase-breaking scattering respectively. Their treatment in a quasi-equilibrium situation is discussed in 
\cite{Szymanska02,Szymanska03}.

The first term in Eq. (\ref{eq:baths}) represents dynamic or static fluctuations of the excitation energy $\epsilon_j$. Provided these fluctuations are slow and weak enough, they are relatively harmless to the ground state: the wavefunction is robust against static disorder in the energy levels, in a similar way that a singlet superconductor is insensitive to weak charge disorder.

The second term is more dangerous, and if a static potential plays just the same role as magnetic impurities in a superconductor\cite{AG,Zittartz}. This corresponds to scattering that breaks up the electron-hole pair (in order for it to be relevant, one must relax the two-level constraint). At the mean field level, this leads first to a reduction in the gap, and then to a gapless excitation spectrum that is still phase coherent. If one is at excitation levels $\rho_x < 1/2$, then with increasing disorder the coherent state is suppressed. But at larger excitation levels $\rho_x > 1/2$, the coherent state remains: with increasing disorder the order parameter becomes dominated by the light field, the excitation spectrum becomes uniform, and the coherent electronic polarisation is continuously reduced to very small values. Such a gapless condensate reproduces the conventional semiconductor laser as an incoherent electron-hole system, with no bound excitons. But this is a very different state than we would have got if we had modelled a high density electron-hole system with Eq. (\ref{varwfn2}) --- such a state has a gap in the spectrum (for example, the region labelled as a ``interacting polariton condensate'' in \figref{fig:phasediag}). One can add pairbreaking scattering to such a state to explore the close formal analogy with a superconductor, though the two-component light-supported order parameter again generates new physical regimes\cite{Marchetti04}. But all in all, incoherent pair-breaking and high electron-hole densities (which tend to go together) drive one into the conventional laser regime. Strong pair-breaking destroys entirely the ``interacting polariton condensate'' regime of \figref{fig:phasediag}.

The last term in Eq. (\ref{eq:baths}) does not exist in a superconductor where it would be forbidden by symmetry. This is an XY-like random-field term (coupling to $S_x$, $S_y$ if we represent two-level systems as a spin model); it is sensitive to the phase of the local order parameter. Such a term will formally destroy the long-range order of the condensate even if infinitesimal (in dimensions below 4) --- but since we have a system with long-range coupling via the optical field, many physical effects of the ground state will remain in the limit of large system size. The role of this term is presumably to suppress the quantum fluctuations in a finite system, and to lead to slow diffusive dynamics of the semiclassical order parameter. But it has not yet been studied carefully. Certainly when this term is large enough, it will be expected to lead us toward the conventional solid state laser model of localised excitations but with rapid dephasing. For small systems, this leads us toward the regime of the ``few-atom laser''\cite{Kimble}, but potentially in the strong coupling regime of large entanglement.

In most practical situations, the effects of scattering and decoherence will strongly suppress the coherent phases appearing at high pumping levels in \figref{fig:phasediag}, and replace them with more conventional weak coupling lasers. Some recent experiments have however demonstrated spontaneous coherence in the regime that can be termed a polariton laser\cite{Deng02}.

\section{Conclusions}

This review has attempted to link the central idea of coherence across the very different physical systems of a dilute Bose gas, excitons, and polaritons. Using a microscopic model of a coherent state wavefunction, and the macroscopic consequence of phase coherence, the many parallels between these systems - and that of superconductivity - are exposed. Furthermore, by understanding the effects of static or dynamic symmetry breaking fields, we can provide a theoretical framework to connect to the classical regime of the laser.

There are many things left out. Because our concern has been with the structure of the theory, we have not discussed experimental systems and experiments except superficially. Nor have we discussed at any length the physical consequences of condensation and hence the critical experimental tests -- though some of these are implicit. 

We have also addressed only thermal equilibrium. Dealing with strongly driven systems that are far from thermal equilibrium is an interesting and difficult challenge that is worth extended theoretical effort. One of the interesting features of the experimental systems is that they are routinely driven very far from equilibrium, into regimes that are impossible to reach in conventional solids. There are many avenues that are yet to be addressed: Can one maintain coherence in a pumped - but perhaps steady state - system? What is the temporal evolution as condensation develops? Non-equilibrium methods using Langevin dynamics, and the language of stimulated scattering, are well developed in the laser arena, and those ideas have been applied to polaritons (see e.g. \cite{Kavokin03}) but it is not known what replaces this approach in the coherent case - as we argued above and elsewhere\cite{Szymanska03} the Langevin equation has no place when phase coherence is dominant.

We have focussed on bulk systems, and in the case of the polariton condensate, nearly mean-field-like systems. For example, the cold fermion systems coupled via a Feshbach resonance have a formally similar theory\cite{Timmermans01} to polaritons to describe them, but however with a mediating boson - in this case a molecule - that is much {\em smaller} than the characteristic separation between fermions. Also the polariton systems are not unconfined (though driven inhomogeneously), but nevertheless have spatial structure that is currently unexplained\cite{Deng03}. A further exciting direction is to small systems with few photons, where the quantum statistics can be exposed. Again this is a regime that is hard to reach in conventional solids, but is quite evident in optics.

\ack
This work is supported by by the EPSRC and by the EU Network ``Photon
mediated phenomena in semiconductor nanostructures''. PRE acknowledges
the financial support of Sidney Sussex college, Cambridge, and the
hospitality of the NHMFL. FMM acknowledges the financial support of
EPSRC (GR/R95951). The NHMFL is supported by the National Science
Foundation, the state of Florida and the US Department of Energy.

\end{document}